\documentclass[a4paper,11pt]{article}

\usepackage[T1]{fontenc}
\usepackage[english]{babel}
\usepackage[utf8]{inputenc}
\usepackage[normalem]{ulem}
\usepackage{comment}

\usepackage[vmargin=2.4cm,hmargin=2.4cm,includeheadfoot]{geometry}
\usepackage[skip,indent,parfill]{parskip}

\usepackage{color}
\usepackage{subcaption}
\usepackage{tikz}
\usetikzlibrary{arrows.meta}
\usetikzlibrary{tikzmark}
\usetikzlibrary{shapes.multipart}
\usetikzlibrary{decorations.markings}

\definecolor{darkblue}{rgb}{0.1,0.1,.7}
\usepackage[colorlinks, linkcolor=darkblue, citecolor=darkblue, urlcolor=darkblue, linktocpage, pagebackref]{hyperref}
\usepackage[dont-mess-around]{fnpct}

\usepackage[numbers,sort&compress]{natbib}

\makeatletter
\patchcmd\NAT@citexnum{\let\NAT@last@num\NAT@num}{\MakeLinkTarget[cite]{}\Hy@backout{\@citeb\@extra@b@citeb}\let\NAT@last@num\NAT@num}{}{\fail}
\makeatother

\usepackage{booktabs}

\usepackage{amsmath}
\usepackage{amssymb}
\usepackage{mathrsfs}
\usepackage{amsthm}
\usepackage{mathtools}
\usepackage{enumitem}
\usepackage{braket}

\usepackage{multirow}

\usepackage{nicematrix}

\numberwithin{equation}{section}

\let\originalleft\left
    \let\originalright\right
\renewcommand{\left}{\mathopen{}\mathclose\bgroup\originalleft}
    \renewcommand{\right}{\aftergroup\egroup\originalright}

\newcommand{\N}{\mathbb{N}}
\newcommand{\Z}{\mathbb{Z}}
\newcommand{\R}{\mathbb{R}}
\newcommand{\id}{\text{\usefont{U}{bbold}{m}{n}1}}

\newcommand{\be}{\begin{equation}} \newcommand{\ee}{\end{equation}}

\renewcommand{\tilde}{\widetilde}

\DeclareMathOperator{\tr}{tr}

\usepackage{xifthen}
\newcommand{\dif}[1][]{\mathord{\ifthenelse{\isempty{#1}}{\mathrm{d}}{\mathrm{d}^{#1}}}}

\begin{document}

\begin{titlepage}
    \vspace*{1em}
    \begin{center}
        {\LARGE \bfseries Thermal effective action for the $O(N)$ vector model} \\[3em]
        {\bf
            Lorenzo Mauro$^{a,b}$,  Alessandro Vichi$^{b}$ 
        } \\[2em]
         \( {}^{a} \) {\itshape Scuola Normale Superiore, Piazza dei Cavalieri 7, I-56126 Pisa, Italy} \\

        \( {}^{b} \) {\itshape Department of Physics, University of Pisa and INFN, \\Largo Pontecorvo 3, I-56127 Pisa, Italy} \\[2em]
        {\small{\texttt{alessandro.vichi@unipi.it}}}
    \end{center}
    \vspace{3em}
    \begin{abstract}
We compute the leading coefficients of the thermal effective action for the critical $O(N)$ vector model in three dimensions, in the large-$N$ limit in presence of non vanishing angular twist. At high temperature, the partition function on a product of a two-dimensional spatial manifold and a thermal circle admits a Kaluza-Klein reduction to a local effective action on the spatial slice, whose coefficients encode universal CFT data such as the Casimir energy and the response to a Kaluza-Klein gauge field. We determine these coefficients through two independent computations: an evaluation of the twisted partition function on the two-sphere in the high-temperature limit, and a direct path-integral computation on a generic weakly curved background. 
The two methods yield consistent results, providing a non-trivial check of the thermal effective action framework.
\end{abstract}
\end{titlepage}

\tableofcontents

\clearpage

\section{Introduction}\label{sec:intro}

Conformal field theories (CFTs) at finite temperature provide a rich playground for exploring the interplay between symmetry, dynamics, and thermal physics. In two dimensions, modular invariance of the torus partition function leads to Cardy's asymptotic formula for the density of states~\cite{Cardy:1986ie}, while finite-size and finite-temperature corrections relate the free energy to the central charge~\cite{Blote:1986en,Affleck:1986bv}. While the zero-temperature properties of CFTs have been extensively studied through the powerful machinery of the conformal bootstrap~\cite{Rattazzi:2008pe,Poland:2018epd} and other analytical techniques, finite-temperature observables present unique challenges. The introduction of a temperature scale breaks conformal invariance explicitly, yet the underlying CFT structure continues to constrain the form of physical quantities, often in subtle and powerful ways~\cite{Iliesiu:2018fao,Gobeil:2018fzy,Petkou:2018ynm,Manenti:2019wxs,Marchetto:2023xap,Buric:2024kxo,Buric:2025uqt,Barrat:2024aoa}.

A particularly fruitful approach to thermal CFTs is the construction of a \emph{thermal effective action}, which captures the universal behavior of the theory in the high-temperature regime $\beta/r \ll 1$, where $\beta$ is the inverse temperature and $r$ is a characteristic length scale of the spatial manifold. In this limit, the compact thermal circle $S^1_\beta$ becomes much smaller than any other scale in the problem, and one can perform a Kaluza-Klein dimensional reduction. The resulting effective theory lives on the spatial slice $M$ and is organized as a derivative expansion, with coefficients that encode genuine CFT data such as the Casimir energy and transport-related quantities. This philosophy is closely related to the construction of equilibrium partition functions in hydrodynamics~\cite{Banerjee:2012iz,Jensen:2012jh}, where locality and symmetry on the thermal manifold dictate the form of allowed terms. Related high-temperature and asymptotic approaches include higher-dimensional Cardy-like formulas for CFTs on spatial tori~\cite{Shaghoulian:2015kta}. The framework relevant to the present work, developed in~\cite{KKCFT}, provides a systematic way to extract universal predictions from any thermal CFT, independent of the microscopic details of the theory; recent applications and refinements include~\cite{Benjamin:2024kdv}.

The coefficients appearing in the thermal effective action are theory-dependent and constitute physical observables of the underlying CFT. Computing them in concrete examples is therefore an important task, both as a check of the formalism and as a way to generate new CFT data. Among the various models accessible to direct computation, the critical $O(N)$ vector model in three dimensions stands out as a paradigmatic example, both because of its physical relevance to a wide class of critical phenomena and because of its role as a testing ground for bootstrap~\cite{Kos:2013tga,Kos:2015mba} and holographic~\cite{Klebanov:2002ja,Giombi:2012ms} approaches to CFT. In the limit $N \to \infty$ with the quartic coupling sent to its critical value, the theory becomes solvable via the Hubbard-Stratonovich transformation, which trades the original interaction for a Gaussian path integral coupled to an auxiliary scalar field $\sigma$. The remaining functional integral can then be evaluated by saddle point, and many physical quantities reduce to the computation of functional determinants on the relevant background. The large-$N$ solution of vector models and the relation between quartic vector models and nonlinear sigma models are reviewed in~\cite{Moshe:2003xn}. This approach has a long history in the study of the model at finite temperature and has been revisited in a number of recent works~\cite{Diatlyk:2023msc,David:2023uya,David:2024pir,david2025largenvectormodel}. These works computed thermal one-point functions, free energies, and related observables in various geometries and limits.

In this paper, we compute the leading coefficients of the thermal effective action for the critical $O(N)$ model on $M \times S^1_\beta$, where $M$ is a two-dimensional Riemannian manifold. We approach the problem from two complementary angles. First, we evaluate the twisted partition function $Z[\beta,\Omega] = \tr e^{-\beta H + i\beta \Omega J}$ on $S^2_r \times S^1_\beta$, where $J$ generates rotations of the sphere. The free energy on this geometry, in the absence of a twist, was recently computed in~\cite{david2025largenvectormodel} in the high-temperature limit, and we use those results as a starting point. Comparing the small-$\beta$ expansion of the twisted partition function with the prediction from the effective action of~\cite{KKCFT} allows us to extract the coefficients $f$, $c_1$, and $c_2$ associated with the Casimir energy, the Ricci scalar, and the Kaluza-Klein gauge field strength, respectively. Second, we perform a direct computation of the effective action on a generic perturbed flat background $G_{\mu\nu} = \delta_{\mu\nu} + H_{\mu\nu}$, integrating out the fundamental field $\phi_i$ and the auxiliary field $\sigma$ to obtain the EFT coefficients without reference to any specific geometry.

A key technical point that arises in our analysis concerns the saddle-point value of the auxiliary field $\sigma$. On a generic background, $\sigma$ may depend on the spatial coordinates, which would significantly complicate the evaluation of the functional determinant—an issue that has played a non-trivial role in related contexts~\cite{Chai:2020zgq,Diatlyk:2023msc}. However, by exploiting general results on perturbed optimization problems, collected in the appendix, we show that at the order considered one may treat $\sigma$ as a constant. This drastically simplifies the calculation and allows us to obtain closed-form expressions for the coefficients.

Our main results are the explicit values
\begin{equation}
f = \frac{2\zeta(3)}{5\pi}N, \qquad c_1 = 0, \qquad c_2 = -\frac{\sqrt{5}\log\varphi_g}{96\pi}N,
\end{equation}
where $\varphi_g = (1+\sqrt{5})/2$ is the golden ratio. Reassuringly, the two independent computations yield identical answers, providing a non-trivial cross-check of both the effective action formalism and our direct evaluation of the path integral.

The paper is organized as follows. In Section~\ref{sec:setup} we set up the formalism of thermal CFTs and review the construction of the thermal effective action via Kaluza-Klein reduction. In Section~\ref{sec:ON} we specialize to the $O(N)$ model, recall the Hubbard-Stratonovich transformation, and carry out the two computations described above: first the twisted partition function on $S^2_r$, then the direct evaluation of the EFT coefficients on a perturbed flat background. We collect technical material on integral identities and perturbed optimization in the appendices.

\section{Setup} \label{sec:setup}

In this section we introduce the basic framework that will be used throughout the paper. Our presentation closely follows the discussion of \cite{KKCFT}, to which we refer the reader for further details and motivation.

\subsection{Thermal CFTs}

We begin by establishing our conventions and notation for conformal field theories at finite temperature, with a possible chemical potential, or angular potential, for a continuous symmetry.

We start from a given CFT on the Lorentzian manifold $M \times \R$, where $\R$ plays the role of time, and $M$ is assumed to be a 2-dimensional Riemannian manifold. In particular, if $t$ is the time coordinate, and $x_1,x_2$ are coordinates on (a patch of) $M$, then the metric is assumed to take the form:
\begin{equation}
ds^2=-dt^2+g_{ij}dx^i dx^j,
\end{equation}
where the indices $i,j$ run over $1,2$. Now let $Q$ be a generator of a one-parameter continuous group of isometries of $M$ (or rather, its representation on the Hilbert space of the theory). Our goal is to compute expressions of the form:
\begin{equation}
Z_{CFT}[\beta,\alpha]:=\tr e^{-\beta H+i\beta \alpha Q},
\end{equation}
with $H$ the Hamiltonian of the theory, namely the generator of time translations. The parameter $\alpha$ plays the role of a chemical potential conjugate to the charge $Q$, while $\beta$ is the inverse temperature.

To obtain a path integral representation of $Z_{CFT}[\beta,\alpha]$, recall that a one-parameter group of diffeomorphisms of $M$ determines a vector field on $M$, and that a vector field on a manifold is naturally associated with a differential operator on scalar-valued functions on $M$ \footnote{In fact, in most treatments of smooth manifolds, tangent vectors are defined to be differential operators at a point.}. Let $D_Q$ denote the differential operator associated with this symmetry. Then $Z_{CFT}$ can be written as a path integral on the Euclidean manifold $M \times S^1_\beta$ of $e^{-S_Q}$, where, for scalar fields, $S_Q$ is obtained from the Euclidean action $S_{CFT}$ by replacing every occurrence of $\dot{\phi}$ with $\dot{\phi}+\alpha D_Q \phi$. Here the Euclidean time coordinate $\tau$ runs from $0$ to $\beta$, and differentiation with respect to it is denoted with the dot notation.

This prescription takes a particularly simple form for theories with a standard kinetic term. If the Lagrangian of the theory is of the form

\begin{equation}
\mathcal{L}=-\frac 12 \partial_\mu \phi_i \partial^\mu \phi_i-V(\phi),
\end{equation}

then one can verify that the only effect of $\alpha$ is a modification of the metric on $M \times S^1_\beta$. The new metric is always independent of $\tau$, so that the problem reduces to studying the CFT on a stationary background.

\subsection{Thermal effective action}

We now turn to the thermal effective action that controls the high-temperature behavior of the partition function. The key observation is that, in the limit $\beta \to 0$, the thermal circle becomes much smaller than any other length scale, and one can perform a Kaluza-Klein reduction.

Consider a metric $G$ on $M \times S^1_\beta$, which is independent of the $S^1_\beta$ coordinate. Our goal is to compute
\begin{equation}
Z_{CFT}[G]:=\int \mathcal{D}\varphi\, e^{-S_{CFT}[G,\varphi]},
\end{equation}
in the limit $\beta \rightarrow 0$. Since we have a Kaluza-Klein setup, we can expect a dimensional reduction. More precisely, if we write the metric in Kaluza-Klein form,
\begin{equation}
ds^2=G_{\mu \nu}dx^\mu dx^\nu=e^{2\phi(x)}[g_{ij}(x)dx^idx^j+\beta^2(d\tau+A_i(x)dx^i)(d\tau+A_j(x)dx^j)],
\end{equation}
then we expect that
\begin{equation}
Z_{CFT}[G]=e^{-S_{eff}[\phi,A,g]},
\end{equation}
where $S_{eff}$ is a local action in $\phi,A,g$ that is gauge invariant and diffeomorphism invariant. However, we can do much better than that. Since we are working with a CFT, we are free to perform Weyl transformations \footnote{There is a possible caveat: the transformation may be anomalous. However, since the full spacetime is odd-dimensional, there is no local Weyl anomaly, so we will ignore it.}, and in particular we can remove the dilaton $\phi$. After this Weyl rescaling, we can immediately expand the action in a derivative expansion:
\begin{equation}\label{eq:Seff}
S_{eff}[g,A]=\int_M d^2x \sqrt g \left(-f+c_1 R+c_2 F_{ij}F^{ij}+\ldots \right),
\end{equation}
with $f,c_1,c_2,\ldots$ a set of theory-dependent coefficients, and the dots denoting terms with more derivatives. In particular, $f$ has the interpretation of a Casimir energy density, $c_1$ controls the response to the spatial curvature of $M$, and $c_2$ governs the coupling to the Kaluza-Klein gauge field strength $F_{ij} = \partial_i A_j - \partial_j A_i$. The remainder of this paper is devoted to the explicit computation of these coefficients in the critical $O(N)$ vector model.

\section{$O(N)$ model at finite temperature}\label{sec:ON}

In this section, we consider the $O(N)$ model, defined by the Lagrangian
\begin{equation}\label{eq:ON_lagrangian}
\mathcal{L}=\frac{1}{2}\partial^\mu \phi^i \partial_\mu \phi_i+\frac{\lambda}{2N} (\phi_i\phi^i)^2,
\end{equation}
with $\phi \in \mathbb{R}^N$. We work at the critical point, obtained by taking $\lambda \rightarrow \infty$, and then consider the large-$N$ limit. The theory is placed on $M \times S^1_\beta$, equipped with a Riemannian metric of Kaluza-Klein form. We will also specifically compute the twisted partition function on the sphere, corresponding to the special case $M=S^2_r$ with metric
\begin{equation}\label{eq:twisted_metric}
ds^2=r^2(d\theta^2+\sin^2 \theta (d\varphi+\Omega d\tau)^2)+d\tau^2,
\end{equation}
where $(\theta,\varphi)$ are the usual polar coordinates on the sphere and $\tau \in [0,\beta]$ is a periodic coordinate for $S^1$.

It is worthwhile to review the Hubbard-Stratonovich transformation here, since it allows us to reduce path integrals in the $O(N)$ theory to Gaussian path integrals, which we are able to compute. Recall that we are trying to compute integrals of the form
\begin{equation}
\int \mathcal{D}\phi_i\ \exp\left(-\int d^3 x \sqrt g \left(\frac 12 \phi_i A \phi_i +\frac{\lambda}{2N} (\phi_i\phi^i)^2 \right) \right),
\end{equation}
where $A$ is a symmetric differential operator. To compute such integrals, we rewrite the quartic term as a Gaussian integral over an auxiliary field $\sigma$:
\begin{equation}
\exp \left( -\int d^3 x \sqrt g \frac{\lambda}{2N} (\phi_i \phi^i)^2 \right) = \int \mathcal{D}\sigma \exp \left(\int d^3 x \sqrt g \left(\frac{N}{8\lambda} \sigma^2-\frac 12 \sigma \phi_i \phi^i \right)\right),
\end{equation}
up to an overall normalization constant. Now $\phi$ appears at most quadratically, so we can integrate it out, leaving us with a functional determinant:
\begin{align}
\int \mathcal{D}\phi_i& \exp\left(-\int d^3 x \sqrt g  \left(\phi_i O \phi_i +\frac{\lambda}{2N} (\phi_i\phi^i)^2 \right) \right) \nonumber\\
&= \int \mathcal{D}\phi_i \mathcal{D}\sigma \exp \left(-\int d^3x \sqrt g \left(\frac 12 \phi_i (O+\sigma) \phi_i-\frac{N}{8\lambda} \sigma^2 \right) \right) \nonumber\\
&=\int \mathcal{D}\sigma \det[O+\sigma]^{-\frac N2} \exp \left(\int d^3 x \sqrt g \frac N {8\lambda} \sigma^2 \right) \nonumber\\
&=\int \mathcal{D} \sigma \exp \left(-\frac N2  \left(\tr \log(O+\sigma)-\int d^3 x \sqrt g\frac N {8\lambda} \sigma^2 \right) \right),
\end{align}
where the power $N$ of the determinant arises from the fact that there are $N$ integrals over $\phi$. Since we need to take the limit $\lambda \rightarrow \infty$, we are left with
\begin{equation}\label{eq:saddle_integral}
\int \mathcal{D} \sigma \exp \left(-\frac N2  \tr \log(O+\sigma) \right),
\end{equation}
which we can evaluate using the saddle-point method, since there is a factor of $N$ multiplying the exponent. The upshot is that we determine $\sigma$ by extremizing the large-$N$ effective action. This trace is generically hard to compute, unless $\sigma$ happens to be uniform. If we have enough symmetries, we will be able to argue that the saddle point is uniform, which will greatly simplify our calculations.

Before starting explicit computations, let us summarize the result we expect, based on the form of the effective action. The following expression follows from the effective action of~\cite{KKCFT}, after restoring the radius $r$\footnote{In~\cite{KKCFT} the radius is set to $r=1$, and the result is written in general dimension.}:
\begin{equation}\label{eq:Z_expansion}
\log Z[\beta,\Omega]=-\frac{4\pi r^2}{1+(r\Omega)^2}\left(-fT^2+\frac{2}{r^2}c_1+\left(2c_1+\frac{8}{3} c_2\right)\Omega^2+\ldots\right).
\end{equation}

This was obtained simply by computing $\int_{S^2_r} d^2x \sqrt g\, \mathcal{O}(x)$, where $\mathcal{O}(x)=1, R, F_{ij}F^{ij}$ in turn. Recall that these were defined on the metric that was rescaled to eliminate the dilaton.

\subsection{Warm-up: $\Omega=0$ on $S^2$}

To begin, let us review the computation of $Z[\beta]$ on $S^1_\beta \times S^2_r$. The results obtained here will be needed later. In this case, the operator $O$ is just (the negative of) the Laplacian on $S^1_\beta \times S^2_r$, which has eigenvalues $\omega_n^2+\frac{l(l+1)}{r^2}$, with $\omega_n=\frac{2\pi n}{\beta}$.

Since the setup is invariant under rotations of the sphere and shifts in the $S^1_\beta$ coordinate, by symmetry we can conclude that the saddle point is uniform. Hence
\begin{equation}
\log Z[\beta]=-\frac N2 \sum_{n \in \Z} \sum_{l=0}^\infty (2l+1)\log \left( \omega_n^2+\frac{l(l+1)}{r^2}+\sigma_0 \right),
\end{equation}
where $\sigma_0$ is the saddle-point value.

This sum has been computed in \cite{david2025largenvectormodel} in the limit we are interested in, namely $\beta/r\ll 1$. The result is
\begin{equation}\label{eq:logZ_untwisted}
\frac 1N \log Z[\beta]=\frac{4\pi r^2 }{\beta^2} \left[\frac{2\zeta(3)}{5\pi}+O\left(\frac{\beta^4}{r^4} \right) \right].
\end{equation}

In particular, note the absence of terms of order $O\left((\beta/r)^0 \right)$. In fact, $\sigma_0$ has no terms of order $1/(\beta r)$.

\subsection{Leading corrections for $\Omega\neq0$}

Our goal in this subsection is to compute the terms
of order $(r\Omega)^2$ and
$(r\Omega)^2(\beta/r)^2$
in the high-temperature expansion of
$\log Z[\beta,\Omega]$.

Here, the differential operator is $O=-\nabla^2_{S^2_r}-(\partial_\tau-\Omega \partial_\varphi)^2$, with $\tau$ the $S^1_\beta$ coordinate ranging from $0$ to $\beta$, and $\varphi$ the longitude on the sphere. The symmetry group is reduced compared to the previous subsection, so the saddle-point value of $\sigma$ might depend on the latitude in the exact problem. For the purposes of the present calculation, this dependence does not contribute at the order of interest.

To see this, we invoke the results of Appendix~\ref{POP} and Appendix~\ref{POP2}. There, we assume that the function we want to extremize is jointly analytic in the variable over which we want to extremize ($\sigma$ in our case) and in two additional parameters ($\beta^2, \Omega^2$). The calculation of the term of order $(r\Omega)^2$ is just a perturbed optimization problem in one parameter, namely $\Omega^2$, since we can simply set $\beta=0$. The calculation we need to perform is of first order in $\Omega^2$. By the results of Appendix~\ref{POP}, we deduce that we do not actually need to compute corrections to $\sigma$, so we can simply treat $\sigma$ as a constant\footnote{Even if such corrections were needed, since we are working with a CFT and set $\beta/r$ to zero, this amounts to sending $r$ to infinity. Then $S^2_r$ becomes $\R^2$, restoring full translational symmetry and making the saddle-point value of $\sigma$ constant again.}.

The computation of the other term seems, on the surface, to be trickier. However, using the notation of Appendix~\ref{POP2}, we are computing $\frac{\partial^2 g}{\partial t \partial u}\big|_{t=u=0}$, which means we only need to find one of the two first-order corrections to $\sigma$. We have already seen above the correction corresponding to $\beta$, which is zero. Therefore, here too we do not need to take into account the fact that the exact saddle-point value of $\sigma$ depends on the latitude.

We still need to check our analyticity conditions, which for us just means checking that the expansion of $\tr \log (O+\sigma)$ contains only even powers of $\beta$ and $\Omega$. To do this, we write
\begin{multline}
\tr \log (O+\sigma)=\tr \log\!\left((-\nabla^2_{S^2_r}+\sigma_0)-(\partial_\tau-\Omega \partial_\varphi)^2+\delta \sigma\right) \\
=\tr \log(-\nabla^2_{S^2_r}+\sigma_0)+\tr \log\!\left(1-(-\nabla^2_{S^2_r}+\sigma_0)^{-1}\bigl((\partial_\tau-\Omega \partial_\varphi)^2-\delta \sigma\bigr)\right),
\end{multline}
where $\sigma=\sigma_0+\delta \sigma$, with $\sigma_0$ the unperturbed saddle-point value of $\sigma$ (which, in particular, is constant). Taylor expanding the last logarithm, one notices that any terms with an odd power of $\Omega$ contain the trace of an odd power of $\partial_\varphi$ times an odd power of $\partial_\tau$ times a spherically symmetric operator. Such a trace vanishes by symmetry. Once these odd terms are seen to be zero, we observe that all remaining terms have powers that are even in both $\beta$ and $\Omega$.

At this point, we can compute $\tr \log(O+\sigma)$ by noting that the eigenvalues of $O$ are $\frac{l(l+1)}{r^2}+(\omega_n+\Omega m)^2$ for $n \in \Z,\, l \in \N,\, m=-l,\ldots,l$. We are therefore left with the sum
\begin{equation}
\log Z[\beta,\Omega]=-\frac N2 \sum_{n \in \Z} \sum_{l=0}^\infty \sum_{m=-l}^l \log \left( (\omega_n+\Omega m)^2+\frac{l(l+1)}{r^2}+\sigma \right),
\end{equation}
which is more involved than the untwisted sum. The result, at the order we are interested in, is
\begin{equation}\label{eq:logZ_twisted}
\frac 1N \log Z[\beta,\Omega]=\frac{4\pi r^2 }{\beta^2}\frac{2\zeta(3)}{5\pi}\bigl(1-(r\Omega)^2\bigr)+\frac{\sqrt 5}{9} \log \varphi_g \cdot (r\Omega)^2,
\end{equation}
with $\varphi_g$ denoting the golden ratio here.

From this expression, we can deduce the values of the coefficients $f, c_1, c_2$ of the effective action by comparing with \eqref{eq:Z_expansion}. We find
\begin{align}\label{eq:EFTcoeff}
f=\frac{2\zeta(3)}{5\pi}N, \qquad
c_1=0, \qquad
c_2=-\frac{\log \varphi_g}{96 \pi} \sqrt 5\, N.
\end{align}

In the next subsection, we will rederive these coefficients by integrating out the field $\phi_i$ directly, without assuming any particular background.

\subsection{Calculation of the EFT coefficients}

We now work on $M \times S^1_\beta$. Since we only need local coefficients in the effective action, we may take $M=\mathbb{R}^2$. As a metric, we put $G_{\mu \nu}=\delta_{\mu \nu}+H_{\mu \nu}$, with $H_{\mu \nu}$ small and independent of $\tau$, where $\tau$ is the coordinate of $S^1$ and lies in the interval $[0,\beta]$. However, $H$ will be allowed to depend on $x,y$, the coordinates of $\mathbb{R}^2$. Here and in what follows, the indices $\mu,\nu,\lambda,\rho$ run over $0,1,2$, while $i,j,k,l$ run over $1,2$.

As a reminder, we obtained above that
\begin{equation}
\log Z=-\int_M d^2 x \sqrt g \left(-f+c_1 R+c_2 F_{ij}F^{ij} +\ldots\right),
\end{equation}
where $g_{ij}$ is the restriction of $G_{\mu \nu}$ to $M$. We immediately run into an issue: we cannot compute $c_1$ with this background. This is because the $c_1$ term is multiplied by $R$, which is then integrated over a 2-dimensional manifold. By the Gauss-Bonnet theorem,
\begin{equation}
\int_M R \sqrt g\, d^2 x =4\pi \chi(M),
\end{equation}
where $\chi(M)$ is the Euler characteristic of the manifold, in particular a topological invariant. Therefore $c_1$ will be multiplied by $\chi(M)$ in the final result for $\log Z$, so if we start from a background with $\int R=0$, we cannot extract $c_1$ from this perturbative expansion around flat space. We nevertheless want to start from an unperturbed background with $R=0$, because we want the unperturbed problem to be easy. Thus, we give up on computing $c_1$ in this way, and instead rely on the value extracted from our earlier computation of $Z[\beta]$ on the sphere (note that we are allowed to set $\Omega=0$ for this purpose).

Let us also recall the definition of $Z$:
\begin{equation}
Z=\int \mathcal{D} \phi_i\, e^{-\int_{M \times S^1_\beta} d^3 x \sqrt G \, \mathcal{L} }.
\end{equation}

Recall that, in the $O(N)$ model with $\lambda \phi^4$ interaction in the limit $\lambda \rightarrow \infty$, what we need to do is to extremize the large-$N$ effective action as a functional of $\sigma$, with $\sigma$ a scalar field on the manifold, and
\begin{equation}
-2\log Z[\sigma]:=\tr \log \left( -\nabla_G^2+\sigma \right),
\end{equation}
where $\nabla^2_G$ is the Laplace-Beltrami operator on $M \times S^1$ with metric $G$. Our goal is to compute the terms up to second order in $H$ of $-2\log Z[\sigma_{opt}]$ that do not contain more than 2 derivatives.

We now expand the Laplace-Beltrami operator as a series in $H$. We adopt the convention that indices are raised and lowered with $\delta^{\mu \nu}$ and $\delta_{\mu \nu}$, respectively, rather than with $G$. In particular, $H:=\delta^{\mu \nu}H_{\mu \nu}$. To the order we are interested in,
\begin{align}
-\nabla^2_G&=D_0+D_1+D_2+\ldots,\\
D_0&:=-\partial^2,\\
D_1&:=\partial_\mu H^{\mu \nu}\partial_\nu-\frac{1}{2} H^{,\mu}\partial_\mu,\\
D_2&:=-\partial_\mu H^\mu_\lambda H^{\lambda \nu} \partial_\nu+\frac{1}{2}H^{\mu \nu}H_{,\mu} \partial_\nu+\frac{1}{4} \partial_\mu(H^{\lambda \sigma} H_{\lambda \sigma}) \partial^\mu.
\end{align}

In particular, $D_i$ contains all terms of order $i$ in $H_{\mu \nu}$. Thus, we need to compute and extremize
\begin{equation}
\tr \log(D_0+D_1+D_2+\ldots+\sigma).
\end{equation}

Now let $\sigma=:\sigma_0+ \tilde{\sigma}$, where $\sigma_0$ is the saddle-point solution of the unperturbed problem. This has already been computed and is, in particular, a uniform field. We then have
\begin{multline}
-2\log Z[\sigma]=\tr \log (D_0+D_1+D_2+\ldots+\sigma_0+\tilde{\sigma}) \\
=\tr \Bigl[\log(D_0+\sigma_0)+(D_0+\sigma_0)^{-1}D_1+(D_0+\sigma_0)^{-1} \tilde{\sigma}+(D_0+\sigma_0)^{-1}D_2 \\
-\frac{1}{2}(D_0+\sigma_0)^{-1}D_1(D_0+\sigma_0)^{-1}D_1-\frac{1}{2}(D_0+\sigma_0)^{-1} \tilde{\sigma}(D_0+\sigma_0)^{-1} \tilde{\sigma} \\
- (D_0+\sigma_0)^{-1} D_1 (D_0+\sigma_0)^{-1} \tilde{\sigma}+\ldots\Bigr].
\end{multline}

In Appendix~\ref{POP}, we show that if we want to compute the optimal value of a function at second order in some parameter ($H$), we only need to perform inversions if the mixed second derivative with respect to the parameter and the variable we are optimizing ($\sigma$) is nonzero. In our case, finding that derivative amounts to computing terms that are first order in both $H$ and $\tilde{\sigma}$. We will need to compute many integrals of the form $\sum_n \int \frac{d^2 k_\perp}{(2\pi)^2} f(\omega_n,k_\perp)$, where $k_\perp$ refers to the projection of the vector $k$ onto the $x_1$-$x_2$ plane, so we define
\begin{equation}
\int \tilde{dk}:=\sum_{n} \int \frac{d^2k}{(2\pi)^2},
\end{equation}
where it is understood that $k^\mu$ is a vector in $\R^3$, whose $0$th component is $k^0=\omega_n$, and whose other two components are integrated over in $\int \frac{d^2 k_\perp}{(2\pi)^2}$. The term we need to ensure vanishes is
\begin{equation}
T_{11}:=-\tr \left[ \tilde{\sigma}(D_0+\sigma_0)^{-1} D_1 (D_0+\sigma_0)^{-1} \right].
\end{equation}
We compute this trace by expanding in a momentum basis,
\begin{equation}
\braket{x | k}:=\frac{1}{\sqrt{\beta}} e^{i k \cdot x},
\end{equation}
noting that this is a complete basis satisfying
\begin{align}
\int \tilde{dk} \ket{k}\bra{k}&=\id, \\
\braket{k|k'}&=\delta(k-k'),
\end{align}
where $\delta(k-k')$ denotes a product of the Kronecker delta in the $0$th components and a Dirac delta in the other two. Note also that here the bras and the kets do \textit{not} represent physical states; rather, they represent vectors in the (rigged) Hilbert space of functions on $M \times S^1_\beta$. We also define
\begin{equation}
\epsilon_k(x):=e^{ik \cdot x}.
\end{equation}

Before continuing, we introduce the (inverse) Fourier transforms of $\tilde{\sigma}(x)$ and $H_{\mu \nu}(x)$:
\begin{align}
\tilde{\sigma}(x)&=:\int \tilde{dk}\, \tilde{\sigma}(k) \epsilon_k(x), \\
H_{\mu \nu}(x)&=:\int \tilde{dk}\, H_{\mu \nu}(k) \epsilon_k(x).
\end{align}

We choose not to notationally distinguish between a function and its Fourier transform, leaving the distinction implicit in the choice of argument (or in the context). Our assumption that $H_{\mu\nu}$ is independent of $\tau$ implies that $H_{\mu \nu}(k)=0$ unless $k^0=0$. This will be important shortly.

We are now ready to expand our trace as a set of integrals:
\begin{multline}
T_{11}=-\tr \left[ \tilde{\sigma}(D_0+\sigma_0)^{-1} D_1 (D_0+\sigma_0)^{-1} \right]=-\tr \left[ \tilde{\sigma}(D_0+\sigma_0)^{-1} D_1 (D_0+\sigma_0)^{-1} \id \right] \\
=-\int \tilde{dk}\,\tr \left[ \tilde{\sigma}(D_0+\sigma_0)^{-1} D_1 (D_0+\sigma_0)^{-1} \ket{k} \bra{k} \right] \\
=-\int \tilde{dk}\, \bra{k}\tilde{\sigma}(D_0+\sigma_0)^{-1} D_1 (D_0+\sigma_0)^{-1} \ket{k}, 
\end{multline}
where we have used the cyclicity of the trace. To make further progress, we expand the definition of $D_1$ and insert the Fourier expansions of $\tilde{\sigma}$ and $H_{\mu \nu}$:
\begin{multline}
T_{11}=-\int \tilde{dk} \int \tilde{dq}\int \tilde{dq'}\, \bra{k} \tilde{\sigma}(q') \epsilon_{q'} (D_0+\sigma_0)^{-1} \\
\times \left(\partial_\mu H^{\mu \nu}(q) \epsilon_q \partial_\nu- \frac{i}{2} q^\mu H(q) \epsilon_q \partial_\mu \right)(D_0+\sigma_0)^{-1} \ket{k},
\end{multline}
where $\epsilon_q$ and $\epsilon_{q'}$ denote multiplication operators by those functions. They satisfy
\begin{equation}
\epsilon_q \ket{k}=\ket{q+k},
\end{equation}
and we also have
\begin{align}
\partial_\mu \ket{k}&=ik_\mu \ket{k}, \\
(D_0+\sigma_0)^{-1} \ket{k}&=\frac{1}{k^2+\sigma_0} \ket{k}.
\end{align}
Using these, we obtain
\begin{multline}
T_{11}=-\int \tilde{dk} \int \tilde{dq} \int \tilde{dq'}\, \braket{k | k+q+q'} \tilde{\sigma}(q') \frac{1}{(k+q)^2+\sigma_0} \\
\times \left(i (k_\mu+q_\mu)H^{\mu \nu}(q) i k_\nu-\frac{i}{2}H_{,\mu}(q) k^\mu \right)\frac{1}{k^2+\sigma_0} \\
=\int \tilde{dk} \int \tilde{dq} \int \tilde{dq'}\, \delta(q+q') \tilde{\sigma}(q') \frac{1}{(k+q)^2+\sigma_0} \\
\times \left(i (k_\mu+q_\mu)H^{\mu \nu}(q) i k_\nu-\frac{i}{2}H_{,\mu}(q) k^\mu \right)\frac{1}{k^2+\sigma_0} \\
=\int \tilde{dk} \int \tilde{dq}\, \tilde{\sigma}(-q) \frac{1}{(k+q)^2+\sigma_0}\left(i (k_\mu+q_\mu)H^{\mu \nu}(q) i k_\nu-\frac{i}{2}H_{,\mu}(q) k^\mu \right)\frac{1}{k^2+\sigma_0}.
\end{multline}

At this point, we would like to carry out the integral in $k$. We can hope to do this using Feynman parametrizations, which are standard in similar situations on affine spaces. However, in order to use the standard procedure, we would need to shift $k \mapsto k-xq$, which might not be legal because the $0$th component of the momentum is really discrete and $x \in [0,1]$. Fortunately, $q$ appears in the argument of the Fourier transform of $H_{\mu \nu}$, which (as argued above) vanishes unless $q^0=0$. So this transformation is legal after all. After standard Feynman-parameter manipulations, we find

\begin{equation}\label{eq:T11_final}
T_{11}=\int \tilde{dq}\, \tilde{\sigma}(-q)\int_0^1 dx \int \tilde{dk}\, \frac{(k_\mu k_\nu-x(1-x)q_\mu q_\nu)H^{\mu \nu}(q)-\frac{ix}{2}H_{,\mu}(q)q^\mu}{[k^2+\sigma_0+x(1-x)q^2]^2}.
\end{equation}

Although $T_{11}$ is not identically zero, by virtue of the form of the corrections computed in Appendix~\ref{POP}, we only need the $q=0$ part of the integral above to vanish, since we are interested in corrections with at most two derivatives, i.e. at most two powers of momentum. It remains to check whether
\begin{equation}\label{eq:T_check}
T:=\int \tilde{dk}\, \frac{k_\mu k_\nu H^{\mu \nu}(0)}{(k^2+\sigma_0)^2}\overset{?}{=}0.
\end{equation}

To do this, we finally need to decompose the components of $H_{\mu \nu}$. We work in a conformal frame in which $H_{00}=0$, and we keep the $A_iA_j$ term because it contributes at second order:
\begin{equation}\label{eq:H_decomposition}
H_{\mu \nu}dx^\mu dx^\nu=(h_{ij}+A_i A_j) dx^i dx^j+2A_i dx^i d\tau.
\end{equation}

With this decomposition, in $T$, the only terms that do not immediately vanish by symmetry are those with $\mu=\nu$. If $\mu=\nu=0$, these are multiplied by $H^{00}=0$. So our $k_\mu k_\nu$ term reduces to a $k_i k_j$ term, and this integral is just zero. This is not by coincidence: we can replace $k_i k_j \mapsto \frac{1}{2} k_\perp^2$ by rotational symmetry, so our integral is just a multiple of what we called $I(0,1,2;M_0)$ in Appendix~\ref{Integrals}, where we always take $d=2$ and $\sigma_0=M_0^2$. Using the relations in Appendix~\ref{IntRels}, we find that
\begin{equation}
\frac{\partial}{\partial M} J(M)\Big|_{M_0}=\frac{\partial}{ \partial M}I(0,1,1;M)\Big|_{M_0}=-2M_0 I(0,1,2;M_0).
\end{equation}

Up to an overall factor, $\tr \log (D_0+M^2)$ is given by $J(M)$, and $M_0$ is exactly the value that extremizes $J(M)$; this is the optimal point of the unperturbed problem. Thus the expression above vanishes, and so does our integral.

We are now left to compute the other corrections, and we are free to set $\tilde{\sigma}=0$. These corrections can be computed using exactly the same methods, and we eventually find:
\begin{multline}
\frac{2}{N} S_{eff}(g,A)=-\frac{2}{N} \log Z= \tr\log(-\nabla_G^2+\sigma^2) \\
=-J(M_0) \int d^2x-\frac{1}{2} J(M_0) \int h\, d^2x-\frac{1}{2} J(M_0) \int d^2x\, A_iA^i \\
-\frac{1}{2} J(M_0) \int d^2x\, A_iA^i+ \frac{1}{2} J(M_0) \int d^2x\, h_{ij}h^{ij}-\frac{1}{8} J(M_0) \int d^2x\, h^2 \\
- \frac{1}{4} J(M_0) \int d^2x\, h^{ij}h_{ij}+J(M_0)\int d^2x\, A_iA^i \\
+ \frac{1}{12} I(1,0,2;M_0) \int d^2x\, F_{ij}F^{ij}+\ldots \\
=\int d^2x \left(-J(M_0)\left(1+\frac{1}{2} h - \frac{1}{4} h_{ij}h^{ij} + \frac{1}{8} h^2+ \ldots \right)+ \frac{1}{12}I(1,0,2;M_0)F_{ij}F^{ij}+\ldots \right).
\end{multline}
The above expression matches the expected form, since
\begin{equation}
\sqrt g=1+\frac{1}{2} h - \frac{1}{4} h_{ij}h^{ij} + \frac{1}{8} h^2+ \ldots
\end{equation}

In particular, we obtain
\begin{align}
f=\frac{1}{2}J(M_0)N, \qquad
c_2= \frac{1}{24}I(1,0,2;M_0)N.
\end{align}

Using $J(M_0)=I(0,1,1;M_0)$ (derived in Appendix~\ref{IntRels}), the known value of $M_0$, and the master formula in Appendix~\ref{Integrals}, we finally find
\begin{align}
f=\frac{2\zeta(3)}{5\pi}N, \qquad
c_2=-\frac{\log \varphi_g}{96 \pi} \sqrt 5\, N,
\end{align}
which match the values obtained in \eqref{eq:EFTcoeff}.

\section{Conclusions}

In this work, we have computed the leading coefficients of the thermal effective action for the critical $O(N)$ vector model in three dimensions, in the large-$N$ limit. By performing two independent calculations—one based on the twisted partition function on $S^2_r \times S^1_\beta$ and one based on a direct evaluation of the path integral on a generic perturbed flat background—we have obtained consistent values for the Casimir-energy coefficient $f$, the curvature coefficient $c_1$, and the gauge-field coefficient $c_2$. The agreement between these two approaches provides a non-trivial check of both the thermal effective action framework of~\cite{KKCFT} and the technical machinery we have employed.

A central technical insight of our analysis is that, to the order in derivatives we considered, the saddle-point value of the Hubbard-Stratonovich field $\sigma$ may be treated as constant, even though on a curved background one might expect a non-trivial spatial profile. This simplification rests on the general results on perturbed optimization problems collected in Appendix~\ref{POP}, which show that a mixed second derivative of the optimized value requires only one of the two first-order corrections to the saddle point. In our setting, the relevant first-order correction in $\beta$ vanishes by virtue of the unperturbed saddle-point condition, rendering the explicit dependence of $\sigma$ on the spatial coordinates immaterial. This observation should generalize to a broad class of large-$N$ computations on curved backgrounds and may be useful in other contexts.

The explicit appearance of the golden ratio in $c_2$ is a curious feature of our final answer. In our computation, it arises from the high-temperature expansion of the relevant Bessel-function sums and reflects the specific arithmetic structure of the integrals computed in Appendix~\ref{Integrals}. It would be interesting to understand whether this number has a deeper interpretation, perhaps in terms of the thermal spectrum or the analytic structure of the dimensionally reduced theory.

Several natural extensions of this work suggest themselves. First, one could compute higher-order coefficients in the derivative expansion of the thermal effective action, which would probe more refined features of the CFT and provide additional data for comparison with the thermal bootstrap~\cite{Iliesiu:2018fao,Gobeil:2018fzy,Marchetto:2023xap,Barrat:2024aoa,Buric:2024kxo} and lattice results. Second, our methods extend straightforwardly to other large-$N$ theories—such as the Gross-Neveu model or large-$N$ gauge theories with matter—where similar Hubbard-Stratonovich-like simplifications are available. Third, it would be interesting to study $1/N$ corrections, where the dependence of the saddle-point field on the background geometry can no longer be ignored, and where the techniques developed here would need to be supplemented with a more careful treatment of the fluctuations. Finally, comparing our results with finite-temperature numerical bootstrap bounds, when these become available at the necessary precision, would provide a stringent test of both approaches and, possibly, new constraints on the spectrum of the critical $O(N)$ model.

We hope that the explicit coefficients computed here, together with the methods developed to obtain them, will prove useful in the broader program of understanding finite-temperature physics in conformal field theories.

\paragraph{Acknowledgements:} We would like Domenico Orlando, Francesco Russo and Iliya Buric for interesting  discussions.

\appendix

\section{Integrals}

\subsection*{A Master Formula}  \label{Integrals}

Let us define the integral

\begin{equation}\label{eq:I_def}
I_d(a,b,\alpha;M):=\int \frac{d^dk}{(2\pi)^d} \sum_{n=-\infty}^\infty \frac{\omega_n^{2a}k^{2b}}{(k^2+\omega_n^2+M^2)^\alpha},
\end{equation}

for the values of $\alpha$ for which this converges, and define it by analytic continuation everywhere else. We now derive a representation useful for the high-temperature expansion. To do so, we start from the identity
\begin{equation}
x^{-\alpha}=\frac{1}{\Gamma(\alpha)}\int_0^\infty \frac{dt}{t}\, t^\alpha e^{-xt},
\end{equation}
valid for $\Re(x)>0$ and $\Re(\alpha)>0$. Thus
\begin{multline}
I_d(a,b,\alpha;M)=\int \frac{d^dk}{(2\pi)^d} \sum_{n=-\infty}^\infty \frac{\omega_n^{2a}k^{2b}}{(k^2+\omega_n^2+M^2)^\alpha} \\
=M^{-2\alpha}\int \frac{d^dk}{(2\pi)^d} \sum_{n=-\infty}^\infty \frac{\omega_n^{2a}k^{2b}}{((k/M)^2+(\omega_n/M)^2+1)^\alpha} \\
=\frac{M^{-2\alpha}}{\Gamma(\alpha)}\int_0^\infty \frac{dt}{t}\, t^\alpha e^{-t} \int \frac{d^dk}{(2\pi)^d} \sum_{n=-\infty}^\infty \omega_n^{2a}e^{-\omega_n^2 t/M^2}k^{2b} e^{-k^2t/M^2}.
\end{multline}

The integral in $k$ is easily evaluated in polar coordinates. Performing it, we obtain
\begin{equation}
I_d(a,b,\alpha;M)=\frac{M^{-2\alpha+2b+d-1} \Gamma\!\left(b+\frac{d-1}{2} \right)}{\Gamma\!\left(\frac{d-1}{2} \right)} (4\pi)^{\frac{d-1}{2}} \int_0^\infty \frac{dt}{t}\, t^{\alpha-b-\frac{d-1}{2}} e^{-t} \left(- \frac{4\pi}{\beta} \right)^a \theta^{(a)}\!\left(\frac{4\pi t}{\beta^2 M^2} \right),
\end{equation}
where
\begin{equation}
\theta(x):=\sum_{n=-\infty}^\infty e^{-\pi n^2 x}
\end{equation}
is the Jacobi theta function. It satisfies
\begin{equation}\label{eq:theta_inversion}
\theta(x)=\frac{1}{\sqrt{x}}\theta(1/x),
\end{equation}
as can be seen by Poisson summation. Taking derivatives, one obtains
\begin{equation}\label{eq:theta_derivatives}
\theta^{(n)}(x)=(-1)^n \sum_{m=0}^n T(n,m)\, x^{-1/2-m-n}\, \theta^{(m)}(1/x),
\end{equation}
where the $T(n,m)$ are positive numbers satisfying
\begin{align}
T(0,0)&=1,\qquad T(0,m)=0 \text{ for }m \neq 0,\\
T(n,m)&=T(n-1,m-1)+\left(m+n-\frac{1}{2}\right)T(n-1,m).
\end{align}

These coefficients turn out to be given by
\begin{equation}
T(n,m)=\frac{(2n)!}{(2m)!(n-m)!\, 4^{n-m}}.
\end{equation}

Returning to our integral, applying the substitution $t \rightarrow 1/t$, using the inversion formula \eqref{eq:theta_derivatives} for the derivatives of the theta function, and re-expressing the derivatives as infinite sums, we obtain
\begin{multline}
I_d(a,b,\alpha;M)=\frac{\beta M^{-2\alpha+2a+2b+d}\, \Gamma\!\left(b+\frac{d-1}{2} \right)}{\Gamma\!\left(\frac{d-1}{2} \right)} (4\pi)^{\frac{d}{2}} \sum_{m=0}^a \frac{(2a)!}{(2m)!(a-m)!\,4^{a-m}} \left(\frac{\beta^2 M^2}{4\pi} \right)^m \\
\times \int_0^\infty \frac{dt}{t^{1+\alpha-a-b-m-d/2}}e^{-1/t} \sum_{n \in \mathbb{Z}}(-\pi)^m n^{2m} e^{-\frac{\beta^2 M^2}{4}n^2 t}.
\end{multline}

We now separate the $n=0$ term from the sum, fold the remaining terms into a single sum, and swap the order of summation and integration. The resulting integral is given by a modified Bessel function of the second kind. After all of this, we find
\begin{multline}\label{eq:I_master}
I_d(a,b,\alpha;M)=\frac{\beta M^{-2\alpha+2a+2b+d}\, \Gamma\!\left(b+\frac{d-1}{2} \right)}{(4\pi)^{d/2}\,\Gamma\!\left(\frac{d-1}{2} \right)} \\
\times \Bigg[\frac{(2a-1)!!}{2^a}\Gamma\!\left(\alpha-a-b-\frac{d}{2}\right) \\
+ 4 \sum_{m=0}^a \left(-\frac{\beta^2 M^2}{4} \right)^m \frac{(2a)!}{(2m)!(a-m)!\,4^{a-m}} \sum_{n=1}^\infty n^{2m} \left(\frac{\beta M n}{2} \right)^{\alpha-a-b-m-d/2} K_{\alpha-a-b-m-d/2}(\beta M n)\Bigg].
\end{multline}

This admits an analytic continuation to all values of $\alpha$ except those for which the Gamma function is singular, since $K_\alpha(x)=K_{-\alpha}(x)$ and $K_\alpha(x) \sim \sqrt{\pi/2}\, x^{-\alpha} e^{-x}$ as $x \rightarrow \infty$, with the usual logarithmic behavior when the Bessel index is zero.

\subsection*{Relations among thermal integrals} \label{IntRels}

In this subsection we specialize to three spacetime dimensions, so the spatial momentum is two-dimensional. Recall the definition
\begin{equation}
I(a,b,\alpha;M):=\int \frac{d^2 k}{(2\pi)^2} \sum_{n \in \mathbb{Z}} \frac{\omega_n^{2a} k^{2b}}{(k^2+\omega_n^2+M^2)^\alpha}.
\end{equation}

In polar coordinates,
\begin{equation}
I(a,b,\alpha;M)=\frac{1}{2\pi} \int_0^\infty k\,dk \sum_{n \in \mathbb{Z}}\frac{\omega_n^{2a} k^{2b}}{(k^2+\omega_n^2+M^2)^\alpha}.
\end{equation}

For our first identity, we integrate by parts, by integrating the $k^{2b+1}$ factor and differentiating the rest:
\begin{equation}
I(a,b,\alpha;M)=\frac{1}{2\pi}\, \frac{\alpha}{2b+2}\int_0^\infty k\,dk \sum_n \frac{2k \cdot k \cdot \omega_n^{2a} k^{2b}}{(k^2+\omega_n^2+M^2)^{\alpha+1}}.
\end{equation}

Hence
\begin{equation}\label{eq:Irel_1}
\boxed{I(a,b,\alpha;M)=\frac{\alpha}{b+1}I(a,b+1,\alpha+1;M).}
\end{equation}

For our second identity, we differentiate with respect to $M$:
\begin{equation}\label{eq:Irel_2}
\boxed{\frac{\partial}{\partial M}I(a,b,\alpha;M)=-2\alpha M\, I(a,b,\alpha+1;M).}
\end{equation}

For our third identity, we multiply and divide by $(k^2+\omega_n^2+M^2)$:
\begin{equation}\label{eq:Irel_3}
\boxed{I(a,b,\alpha;M)=I(a+1,b,\alpha+1;M)+I(a,b+1,\alpha+1;M)+M^2 I(a,b,\alpha+1;M).}
\end{equation}

We need one more identity. Define
\begin{equation}
J(M,\alpha):=-\int \frac{d^2 k}{(2\pi)^2} \sum_{n \in \mathbb{Z}} \frac{\log(k^2+\omega_n^2+M^2)}{(k^2+\omega_n^2+M^2)^\alpha},
\end{equation}

via the integral above where it converges, and by analytic continuation otherwise. Applying the same integration-by-parts trick (differentiating the entire fraction and integrating $k$), and then setting $\alpha=0$, we obtain
\begin{equation}\label{eq:J_identity}
\boxed{J(M):=J(M,0)=I(0,1,1;M).}
\end{equation}

\section{Additional technical details}

\subsection*{Perturbed Optimization Problems} \label{POP}

Suppose we want to extremize the function $f(x,t)$ with respect to $x$, and we have managed to find that the optimal value of $x$ at $t=0$ is $0$. We want to study $f(x_{opt}(t),t)$ at second order in $t$.

We begin by expanding
\begin{equation}
x(t)=x_1 t+x_2 t^2+\ldots
\end{equation}
and we want to solve
\begin{equation}
\frac{\partial}{\partial x}f(x,t) = 0.
\end{equation}
To make progress, we Taylor expand $f$ in its two arguments:
\begin{equation}
f(x,t)=f_{00}+f_{10}x+f_{01}t+\frac{1}{2}f_{20}x^2+f_{11}xt+\frac{1}{2}f_{02}t^2+\ldots
\end{equation}
Taking the first derivative with respect to $x$ and setting it to zero,
\begin{equation}
0 \overset{!}{=} f_{10}+f_{20}x+f_{11}t+ \ldots
\end{equation}
Substituting the expansion of $x(t)$ into the above,
\begin{equation}
0 \overset{!}{=} f_{10}+ f_{20}x_1 t+f_{11} t + \ldots
\end{equation}
At orders $0$ and $1$ this gives
\begin{align}
f_{10}=0,\qquad
f_{20}x_1+f_{11}=0.
\end{align}
The first equation simply states that $x=0$ is a solution of the unperturbed problem. The second can be solved for $x_1$:
\begin{equation}
x_1=-f_{20}^{-1} f_{11}.
\end{equation}
For a functional variable, $f_{20}^{-1}$ denotes the inverse Hessian.
We now compute $f(x(t),t)$ up to quadratic order in $t$:
\begin{equation}
f(x(t),t)=f_{00}+f_{01}t+\frac{1}{2}f_{20}(x_1 t+\ldots)^2+f_{11}(x_1 t+\ldots)t+\frac{1}{2}f_{02}t^2+\ldots,
\end{equation}
where we have already used $f_{10}=0$. This would have been the only term carrying a dependence on $x_2$. Thus, to find the optimal value at second order, we only need the optimal point at first order. Under the usual regularity and non-degeneracy assumptions, this is true in general: to find the optimal value at order $n>0$, we only need to find the optimal point at order $n-1$. In particular, at first order, no extra work is required.
Going back to our expression for $f(x(t),t)$,
\begin{multline}
f(x(t),t)=f_{00}+f_{01}t+\frac{1}{2}f_{20}(x_1 t+\ldots)^2+f_{11}(x_1 t+\ldots)t+\frac{1}{2}f_{02}t^2+\ldots\\
=f_{00}+f_{01}t+\frac{1}{2}f_{20} f_{20}^{-1} f_{11} f_{20}^{-1} f_{11} t^2-f_{11} f_{20}^{-1} f_{11} t^2+\frac{1}{2}f_{02}t^2+\ldots\\
=f_{00}+f_{01}t-\frac{1}{2}f_{11} f_{20}^{-1} f_{11} t^2+\frac{1}{2}f_{02}t^2+\ldots
\end{multline}
This is our key result, so let us box it:
\begin{equation}\label{eq:POP_main}
\boxed{f(x(t),t)=f_{00}+f_{01}t-\frac{1}{2}f_{11} f_{20}^{-1} f_{11} t^2+\frac{1}{2}f_{02}t^2+\ldots}
\end{equation}

In particular, we see that we only need to compute $f_{20}^{-1}$ when $f_{11}$ is nonzero.

\subsection*{Two independent variables} \label{POP2}

Suppose we now want to extremize $f(x,t,u)$ with respect to $x$. We are looking for a function $x(t,u)$ such that
\begin{equation}\label{eq:POP2_defining}
f_x(x(t,u),t,u)=0
\end{equation}
for all $t,u$, and we want to compute $g(t,u):=f(x(t,u),t,u)$. We are interested, in particular, in $\frac{\partial^2 g}{\partial t \partial u}\big|_{t=u=0}$. A straightforward computation gives
\begin{align}
g_{tu}(t,u)&=f_x(x,t,u)\, x_{tu}(t,u)+f_{xx}(x,t,u)\, x_t(t,u) x_u(t,u)\nonumber\\
&+f_{xt}(x,t,u)\, x_u(t,u)+f_{xu}(x,t,u)\, x_t(t,u)+f_{tu}(x,t,u).
\end{align}

Differentiating the defining equation \eqref{eq:POP2_defining} with respect to $t$,
\begin{equation}
f_{xx}(x(t,u),t,u)\, x_t(t,u)+f_{xt}(x(t,u),t,u)=0.
\end{equation}

Therefore, using also the defining equation,
\begin{equation}\label{eq:POP2_main}
g_{tu}(t,u)=f_{xu}(x,t,u)\, x_t(t,u)+f_{tu}(x,t,u).
\end{equation}

Note that this is valid everywhere, not just at $t=u=0$. In particular, to compute a mixed second-order correction to an optimal value, we only need one of the two first-order corrections to the optimal point.

\bibliography{refs2}

@article{KKCFT,
    author        = "Benjamin, Nathan and Lee, Jaeha and Ooguri, Hirosi and Simmons-Duffin, David",
    title         = "{Universal asymptotics for high energy CFT data}",
    eprint        = "2306.08031",
    archivePrefix = "arXiv",
    primaryClass  = "hep-th",
    reportNumber  = "CALT-TH-2023-014, IPMU-23-0020",
    doi           = "10.1007/JHEP03(2024)115",
    journal       = "JHEP",
    volume        = "03",
    pages         = "115",
    year          = "2024"
}

@article{david2025largenvectormodel,
    author        = "David, Justin R. and Kumar, Srijan",
    title         = "{The large $N$ vector model on $S^1 \times S^2$}",
    eprint        = "2411.18509",
    archivePrefix = "arXiv",
    primaryClass  = "hep-th",
    doi           = "10.1007/JHEP03(2025)169",
    journal       = "JHEP",
    volume        = "03",
    pages         = "169",
    year          = "2025"
}

@article{Jensen:2012jh,
    author        = "Jensen, Kristan and Kaminski, Matthias and Kovtun, Pavel and Meyer, Rene and Ritz, Adam and Yarom, Amos",
    title         = "{Towards hydrodynamics without an entropy current}",
    eprint        = "1203.3556",
    archivePrefix = "arXiv",
    primaryClass  = "hep-th",
    doi           = "10.1103/PhysRevLett.109.101601",
    journal       = "Phys. Rev. Lett.",
    volume        = "109",
    pages         = "101601",
    year          = "2012"
}

@article{Banerjee:2012iz,
    author        = "Banerjee, Nabamita and Bhattacharya, Jyotirmoy and Bhattacharyya, Sayantani and Jain, Sachin and Minwalla, Shiraz and Sharma, Tarun",
    title         = "{Constraints on Fluid Dynamics from Equilibrium Partition Functions}",
    eprint        = "1203.3544",
    archivePrefix = "arXiv",
    primaryClass  = "hep-th",
    doi           = "10.1007/JHEP09(2012)046",
    journal       = "JHEP",
    volume        = "09",
    pages         = "046",
    year          = "2012"
}

@article{Benjamin:2024kdv,
    author        = "Benjamin, Nathan and Chang, Chi-Ming and Lee, Jaeha and Ooguri, Hirosi and Wang, Yifan",
    title         = "{Angular fractals in thermal QFT}",
    eprint        = "2405.17562",
    archivePrefix = "arXiv",
    primaryClass  = "hep-th",
    doi           = "10.1007/JHEP11(2024)134",
    journal       = "JHEP",
    volume        = "11",
    pages         = "134",
    year          = "2024"
}

@article{Diatlyk:2023msc,
    author        = "Diatlyk, Oleksandr and Popov, Fedor K. and Wang, Yifan",
    title         = "{Beyond $N=\infty$ in large $N$ conformal vector models at finite temperature}",
    eprint        = "2309.02347",
    archivePrefix = "arXiv",
    primaryClass  = "hep-th",
    doi           = "10.1007/JHEP08(2024)219",
    journal       = "JHEP",
    volume        = "08",
    pages         = "219",
    year          = "2024"
}

@article{David:2024pir,
    author        = "David, Justin R. and Kumar, Srijan",
    title         = "{One point functions in large $N$ vector models at finite chemical potential}",
    eprint        = "2406.14490",
    archivePrefix = "arXiv",
    primaryClass  = "hep-th",
    doi           = "10.1007/JHEP01(2025)080",
    journal       = "JHEP",
    volume        = "01",
    pages         = "080",
    year          = "2025"
}

@article{David:2023uya,
    author        = "David, Justin R. and Kumar, Srijan",
    title         = "{Thermal one-point functions: CFT's with fermions, large $d$ and large spin}",
    eprint        = "2307.14847",
    archivePrefix = "arXiv",
    primaryClass  = "hep-th",
    doi           = "10.1007/JHEP10(2023)143",
    journal       = "JHEP",
    volume        = "10",
    pages         = "143",
    year          = "2023"
}

@article{Petkou:2018ynm,
    author        = "Petkou, Anastasios C. and Stergiou, Andreas",
    title         = "{Dynamics of Finite-Temperature Conformal Field Theories from Operator Product Expansion Inversion Formulas}",
    eprint        = "1806.02340",
    archivePrefix = "arXiv",
    primaryClass  = "hep-th",
    doi           = "10.1103/PhysRevLett.121.071602",
    journal       = "Phys. Rev. Lett.",
    volume        = "121",
    number        = "7",
    pages         = "071602",
    year          = "2018"
}

@article{Manenti:2019wxs,
    author        = "Manenti, Andrea",
    title         = "{Thermal CFTs in momentum space}",
    eprint        = "1905.01355",
    archivePrefix = "arXiv",
    primaryClass  = "hep-th",
    doi           = "10.1007/JHEP01(2020)009",
    journal       = "JHEP",
    volume        = "01",
    pages         = "009",
    year          = "2020"
}

@article{Barrat:2024aoa,
    author        = "Barrat, Julien and Marchetto, Enrico and Miscioscia, Alessio and Pomoni, Elli",
    title         = "{Thermal bootstrap for the critical $O(N)$ model}",
    eprint        = "2411.00978",
    archivePrefix = "arXiv",
    primaryClass  = "hep-th",
    doi           = "10.1103/PhysRevLett.134.211604",
    journal       = "Phys. Rev. Lett.",
    volume        = "134",
    number        = "21",
    pages         = "211604",
    year          = "2025"
}

@article{Buric:2024kxo,
    author        = "Buric, Ilija and Russo, Francesco and Schomerus, Volker and Vichi, Alessandro",
    title         = "{Thermal one-point functions and their partial wave decomposition}",
    eprint        = "2408.02747",
    archivePrefix = "arXiv",
    primaryClass  = "hep-th",
    doi           = "10.1007/JHEP12(2024)021",
    journal       = "JHEP",
    volume        = "12",
    pages         = "021",
    year          = "2024"
}

@article{Chai:2020zgq,
    author        = "Chai, Noam and Chaudhuri, Soumyadeep and Choi, Changha and Komargodski, Zohar and Rabinovici, Eliezer and Smolkin, Michael",
    title         = "{Thermal Order in Conformal Theories}",
    eprint        = "2005.03676",
    archivePrefix = "arXiv",
    primaryClass  = "hep-th",
    doi           = "10.1103/PhysRevD.102.065014",
    journal       = "Phys. Rev. D",
    volume        = "102",
    number        = "6",
    pages         = "065014",
    year          = "2020"
}

@article{Klebanov:2002ja,
    author        = "Klebanov, Igor R. and Polyakov, Alexander M.",
    title         = "{AdS dual of the critical O(N) vector model}",
    eprint        = "hep-th/0210114",
    archivePrefix = "arXiv",
    doi           = "10.1016/S0370-2693(02)02980-5",
    journal       = "Phys. Lett. B",
    volume        = "550",
    pages         = "213--219",
    year          = "2002"
}

@article{Giombi:2012ms,
    author        = "Giombi, Simone and Yin, Xi",
    title         = "{The Higher Spin/Vector Model Duality}",
    eprint        = "1208.4036",
    archivePrefix = "arXiv",
    primaryClass  = "hep-th",
    doi           = "10.1088/1751-8113/46/21/214003",
    journal       = "J. Phys. A",
    volume        = "46",
    pages         = "214003",
    year          = "2013"
}

@article{Kos:2013tga,
    author        = "Kos, Filip and Poland, David and Simmons-Duffin, David",
    title         = "{Bootstrapping the O(N) vector models}",
    eprint        = "1307.6856",
    archivePrefix = "arXiv",
    primaryClass  = "hep-th",
    doi           = "10.1007/JHEP06(2014)091",
    journal       = "JHEP",
    volume        = "06",
    pages         = "091",
    year          = "2014"
}

@article{Kos:2015mba,
    author        = "Kos, Filip and Poland, David and Simmons-Duffin, David and Vichi, Alessandro",
    title         = "{Bootstrapping the O(N) Archipelago}",
    eprint        = "1504.07997",
    archivePrefix = "arXiv",
    primaryClass  = "hep-th",
    doi           = "10.1007/JHEP11(2015)106",
    journal       = "JHEP",
    volume        = "11",
    pages         = "106",
    year          = "2015"
}

@article{Cardy:1986ie,
  author = "Cardy, John L.",
  title = "{Operator Content of Two-Dimensional Conformally Invariant Theories}",
  journal = "Nucl. Phys. B",
  volume = "270",
  year = "1986",
  pages = "186--204",
  doi = "10.1016/0550-3213(86)90552-3"
}

@article{Shaghoulian:2015kta,
  author = "Shaghoulian, Edgar",
  title = "{Modular Invariance On $S^1\times S^3$ and Circle Fibrations}",
  journal = "Phys. Rev. D",
  volume = "93",
  year = "2016",
  number = "12",
  pages = "126005",
  eprint = "1508.02728",
  archivePrefix = "arXiv",
  primaryClass = "hep-th",
  doi = "10.1103/PhysRevD.93.126005"
}

@article{Moshe:2003xn,
  author = "Moshe, Moshe and Zinn-Justin, Jean",
  title = "{Quantum field theory in the large N limit: A review}",
  journal = "Phys. Rept.",
  volume = "385",
  year = "2003",
  pages = "69--228",
  eprint = "hep-th/0306133",
  archivePrefix = "arXiv",
  doi = "10.1016/S0370-1573(03)00263-1"
}

@article{Rattazzi:2008pe,
  author = "Rattazzi, Riccardo and Rychkov, Vyacheslav S. and Tonni, Erik and Vichi, Alessandro",
  title = "{Bounding scalar operator dimensions in 4D CFT}",
  journal = "JHEP",
  volume = "12",
  year = "2008",
  pages = "031",
  eprint = "0807.0004",
  archivePrefix = "arXiv",
  primaryClass = "hep-th",
  doi = "10.1088/1126-6708/2008/12/031"
}

@article{Poland:2018epd,
  author = "Poland, David and Rychkov, Slava and Vichi, Alessandro",
  title = "{The Conformal Bootstrap: Theory, Numerical Techniques, and Applications}",
  journal = "Rev. Mod. Phys.",
  volume = "91",
  year = "2019",
  number = "1",
  pages = "015002",
  eprint = "1805.04405",
  archivePrefix = "arXiv",
  primaryClass = "hep-th",
  doi = "10.1103/RevModPhys.91.015002"
}

@article{Buric:2025uqt,
    author = "Buri{\'c}, Ilija and Mangialardi, Francesco and Russo, Francesco and Schomerus, Volker and Vichi, Alessandro",
    title = "{Heavy-heavy-light asymptotics from thermal correlators}",
    eprint = "2506.21671",
    archivePrefix = "arXiv",
    primaryClass = "hep-th",
    doi = "10.1007/JHEP04(2026)027",
    journal = "JHEP",
    volume = "04",
    pages = "027",
    year = "2026"
}

@article{Gobeil:2018fzy,
    author = "Gobeil, Yan and Maloney, Alexander and Ng, Gim Seng and Wu, Jie-qiang",
    title = "{Thermal Conformal Blocks}",
    eprint = "1802.10537",
    archivePrefix = "arXiv",
    primaryClass = "hep-th",
    doi = "10.21468/SciPostPhys.7.2.015",
    journal = "SciPost Phys.",
    volume = "7",
    number = "2",
    pages = "015",
    year = "2019"
}

@article{Iliesiu:2018fao,
    author = "Iliesiu, Luca and Kolo\u{g}lu, Murat and Mahajan, Raghu and Perlmutter, Eric and Simmons-Duffin, David",
    title = "{The Conformal Bootstrap at Finite Temperature}",
    eprint = "1802.10266",
    archivePrefix = "arXiv",
    primaryClass = "hep-th",
    reportNumber = "CALT-TH-2018-013, PUPT-2550",
    doi = "10.1007/JHEP10(2018)070",
    journal = "JHEP",
    volume = "10",
    pages = "070",
    year = "2018"
}

@article{Marchetto:2023xap,
    author = "Marchetto, Enrico and Miscioscia, Alessio and Pomoni, Elli",
    title = "{Sum rules \& Tauberian theorems at finite temperature}",
    eprint = "2312.13030",
    archivePrefix = "arXiv",
    primaryClass = "hep-th",
    reportNumber = "DESY-23-224",
    month = "12",
    year = "2023"
}

@article{Blote:1986en,
    author = "Blote, H. W. J. and Cardy, John L. and Nightingale, M. P.",
    title = "{Conformal Invariance, the Central Charge, and Universal Finite Size Amplitudes at Criticality}",
    journal = "Phys. Rev. Lett.",
    volume = "56",
    pages = "742--745",
    year = "1986",
    doi = "10.1103/PhysRevLett.56.742"
}

@article{Affleck:1986bv,
    author = "Affleck, Ian",
    title = "{Universal Term in the Free Energy at a Critical Point and the Conformal Anomaly}",
    journal = "Phys. Rev. Lett.",
    volume = "56",
    pages = "746--748",
    year = "1986",
    doi = "10.1103/PhysRevLett.56.746"
}

\end{document}